**Ultra-thin yttria-stabilized zirconia as a flexible and stable substrate for infrared nano-optics**

*Kavitha K. Gopalan* [1], *Daniel Rodrigo* [1], *Bruno Paulillo* [1], *Kamal K. Soni* [2] *and Valerio Pruneri* [1,3]*

K.K. Gopalan, Dr. D. Rodrigo, Dr. B. Paulillo, Dr. K.K. Soni, Prof. Dr.V. Pruneri

1. ICFO-Institut de Ciencies Fotoniques,The Barcelona Institute of Science and Technology, 08860 Castelldefels, Spain E-mail: valerio.pruneri@icfo.eu
2. Corning Incorporated, Science & Technology Division Corning, New York, NY 14831, USA
3. ICREA-Institució Catalana de Recerca i Estudis Avançats Passeig Lluís Companys, 23,08010 Barcelona, Spain

Infrared (IR) technology is a rapidly growing field with applications ranging from thermal imaging to chemical and biological IR spectroscopy.[1,2] Several of these applications, especially spectroscopic sensing need substrates that are transparent in the IR. Commonly used transparent substrates, such as fused silica, are transparent only up to about 2.2 $\mu$m. Instead, for longer wavelengths, calcium fluoride ($CaF_2$), magnesium fluoride ($MgF_2$), barium fluoride ($BaF_2$), silicon (Si), germanium (Ge), zinc selenide (ZnSe), can be employed. The majority of these substrates are fragile, hygroscopic or expensive. Moreover, none of them is mechanically flexible, as required by many emerging applications such as curved or bendable IR sensors.[3] When mechanical flexibility is required, polymers, such as polyethylene terephthalate (PET) and polyethylene naphthalate (PEN), are used in the visible region, but cannot be extended to the IR as they have several vibrational absorption fingerprints. There exist polymers, including parylene C, polydimethylsiloxane (PDMS), polyimide, that are more transparent in the IR and have been studied in literature.[4] However, ultraviolet lithography is not always easy on such materials due to their sensitivity to chemicals. In addition, high-resolution electron beam lithography on these polymeric substrates is restricted by their limited thermal and radiation tolerance and non-planar nature. Alternate methods like nano-stencil lithography, self-assembly techniques could be employed to implement sub-micron features on these polymer substrates for use in plasmonic sensors.[4] Besides the difficulty of finding an appropriate lithography method for patterning, the transmission spectra of these IR polymer substrates show still many vibrational fingerprints between 2-10 μm, which prevent their use both in most of the the near-IR (1-5 μm) and mid-IR (5-25 μm) regions.

Yttria-stabilized Zirconia (YSZ) is a ceramic that has received a lot of attention due to its exceptional properties such as high hardness, high dielectric constant, chemical inertness and high ionic conductivity at elevated temperatures.[5] In the powder form it is used to make coatings that are chemically inert and tolerant to mechanical wear and tear, for example in cutting tools, chemical tank linings and dental restorations.[6,7] It is also used as an electrolyte in solid oxide fuel cells (SOFCs).[8] Lately, there is a growing interest in using thin films and microspheres of YSZ for various photonic applications. [9,10] ENrG Inc. has commercialized 20 and 40 $\mu$m thick flexible substrates of 3 mol% YSZ (3YSZ), which has also shown remarkable transparency in the near-IR and mid-IR while being translucent in the visible.[11] In this paper, we propose and demonstrate for the first time that 3YSZ can be an ideal platform to implement next generation flexible IR nano-optic devices, such as plasmonic sensors and polarizers. We also show that it can be combined with graphene to make flexible transparent electrodes for the IR that can be used for cell culture spectroscopy and IR transparent shielding. [12,13]

The Fourier Transform-Infrared Spectroscopy (FTIR) transmission and reflection spectra of one-side-polished ultrathin (20 $\mu$m-thick) 3YSZ substrate are shown in **Figure 1a**. The transmittance is more than 75% in the wavelength range from 2 to 10 $\mu$m. Note that due to a sample thickness that is comparable to mid-infrared wavelengths, transmission and reflection spectra show features resulting from Fabry-Perot interference (continuous curve in Figure 1a). In the same graph, the average transmission/reflection is plotted by taking the spectra at low resolution. From Fig 1(b) we appreciate the translucent aspect of such thin 3YSZ substrate under visible light while the IR camera image in Fig 1(c) highlights the IR transparency of the material (left image): For comparison we also show Corning Willow glass (125 $\mu$m think) which is opaque in the thermal IR range (right image).

The average roughness (RMS value) of the smooth side of the ultra-thin-3YSZ substrate is around 20 nm. This level of roughness is much smaller than the IR wavelengths, so that surface scattering effects are negligible.

Flexible nano-antennas and wire-grid polarizers

In order to demonstrate the potential of ultra-thin 3YSZ as flexible transparent substrate in the mid-IR, we fabricated gold dipole antenna arrays which are widely used as localized surface plasmon resonance (LSPR) based sensors. [14] We used standard double layer polymethyl methacrylate (PMMA) e-beam lithography to define dipole arrays of various dimensions. Subsequently, 100nm Au with an adhesion layer of Ti was deposited and lifted off. Unlike CaF$_2$ or BaF$_2$ substrates, ultra-thin 3YSZ shows no adhesion issues with resists and metals.[15] Contrary to 3YSZ, fluoride substrates are hygroscopic and cannot be used in the long term for sensors in humid or harsh environments.[16] Since the spectral range of operation is related to the antenna geometry and dimension, we fabricated gold dipole arrays of different lengths $L$=1.5 $\mu$m to $L$=2.8 $\mu$m) and periods ($P$=1.4$L$). Figure 2(b) shows the SEM image of a typical gold dipole antenna arrays fabricated on ultra-thin 3YSZ. Bruker FTIR Hyperion microscope was used to measure the reflection and transmission spectra. Measurements were performed with light polarized parallel and perpendicular to the axis of the dipoles. As expected, plasmonic modes are excited only when incident light is polarized parallel to the axis of the dipoles.

To demonstrate mechanical flexibility, the substrate with gold dipole arrays was fixed in between two movable aluminium rails and repeatedly bent. FTIR measurements were carried out in reflection mode while the substrate was at a bend radius, r=2.2 cm. Figure 2 (c) confirms that 3YSZ substrate with dipole antenna can be continuously bent without affecting the optical response. Subsequently, the substrate was subjected to about 100 bending cycles and measured right after. The difference in intensities of the plasmonic response before and after bending is negligible, confirming the capabilities of ultra-thin 3YSZ for use as flexible transparent substrate for IR nano-optics.

Besides its use for plasmonics (nano-optics), ultra-thin 3YSZ substrates offers great potential for other IR devices. For instance, polarizers are widely used devices in the mid-IR. Commercially available holographic wire-grid polarizers use materials like BaF$_2$ , ZnSe, thallium bromoiodide (KRS-5 ) which can be toxic and fragile. These polarizers are also expensive due to complex fabrication techniques. Wire grid polarizers consist of arrays of sub-wavelength metallic wires that transmit radiation with an electric field vector perpendicular to the wire and reflect the radiation with electric field vector parallel to the wires.[17,18] For our experiments, we fabricated by a simple lithographic method Au wire grids with a width of 500 nm and a period of 1500 nm, for working in the 6-10$\mu$m wavelength region. The extinction of our wire grid polarizers reaches up to 15 dB and is comparable to similar reported structures. [19] The demonstrated polariser can potentially be used in next generation flexible IR photonic devices, for example, thermal cameras.

Graphene-based flexible IR transparent conductor and heater

Indium tin oxide (ITO) and silver nanowire (AgNW) films are known to be good transparent conductors and have been used as transparent heaters in the visible region. [20,21] However, they both become highly reflective in the IR. Instead, graphene is known for its high electrical conductivity and low absorption (2.3%) in the visible as well as IR range. As ITO and AgNW are transparent only in the visible range, these films have limited use as a flexible transparent heater. [22] Graphene combined with ultra-thin 3YSZ offers a unique opportunity as IR transparent conductors, especially when one considers that doped graphene can have an absorption lower than 2.3% in the mid-IR, thanks to Pauli blocking effects. [13,23,24] Here we demonstrate a transparent heater but the same graphene on ultra-thin 3YSZ structure can be used for other applications, including electromagnetic (EM) shielding, plasmonics, chemical and biological sensing.

As opposed to flexible polymeric substrates, ultra-thin 3YSZ is planar and rigid and hence standard chemical vapour deposition (CVD) graphene transfer by wet-etching is facilitated. In addition, as it withstands high temperatures, graphene could be grown using mass-scalable techniques over large area, such as CVD. In this paper, graphene initially grown on copper using CVD technique was transferred on several ultra-thin 3YSZ substrates using PMMA as sacrificial layer. The sheet resistance of graphene on ultra-thin 3YSZ was measured to be between 1-1.5 k$\Omega$/sq, which is similar to the value reported on commonly used substrates like Si, SiO2 etc. [25] The corresponding Raman spectrum is shown in **Figure 4a**. The ratio of intensities of the 2D and G peak is greater than 2 indicating good structural quality.

After the electrical and optical characterization, we demonstrated an IR transparent flexible heater. **Figure 4c** shows the thermal image under electrical current. Temperatures over 100 ºC were attained in the experiments for current densities of 1 W/cm$^2$. Given this performance as shown in **Figure 4b,** the proposed IR transparent conductive structure could find applications as ATR (Attenuated Total Reflectance) component in live cell imaging, where maintaining the temperature at physiological conditions (e.g. 37 ˚C) as well as transparency is mandatory.[26–29] In addition it could have great potential in EM shielding windows for detectors.[30]

## Conclusions

In conclusion, we have demonstrated that ultra-thin 3YSZ ceramic is an ideal mechanically flexible platform to implement next generation IR nano-optic devices. In particular, we have combined ultra-thin 3YSZ with metallic nano-structures and graphene to demonstrate plasmonics, polarizers and transparent heaters. The proposed 3YSZ based platform withstands high temperature processing, e.g, direct deposition of graphene, and harsh environments thanks to its high mechanical, thermal and chemical stability. In addition, besides the functional capability of making foldable and bendable devices, the mechanical flexibility of ultra-thin 3YSZ offers the possibility of roll-to-roll processing for low cost and large-scale fabrication processes. Our work points out that ultra-thin 3YSZ is a unique substrate for IR applications which combines multiple features, including mechanical flexibility, durability, transparency, easy processing, which are not available from other available material alternatives.

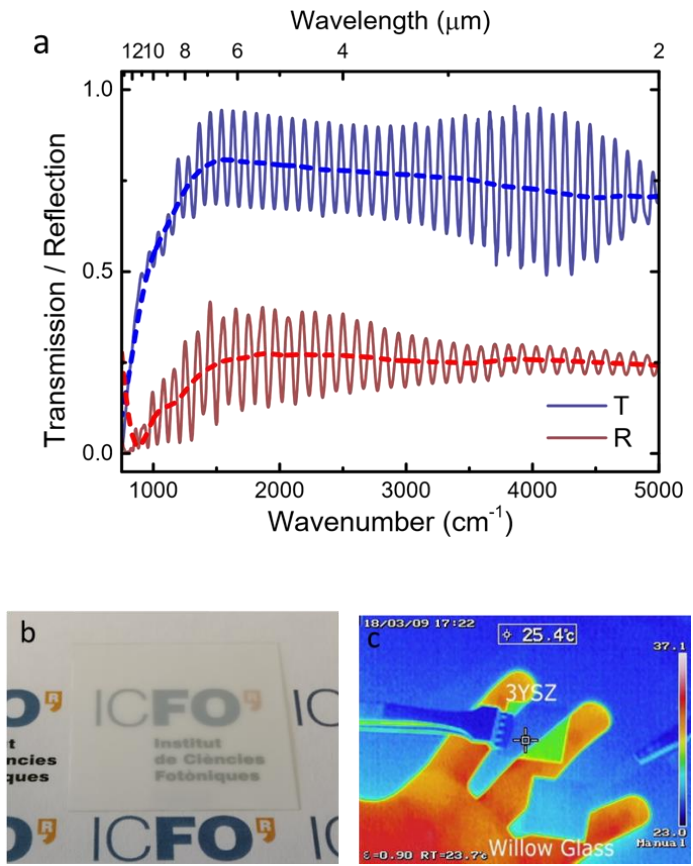

**Figure 1**. ( a) Transmission and reflection spectra of 20 μm thick 3YSZ. (b)Image of ultra-thin 3YSZ substrate that is transparent in the IR and translucent in the visible region. (c)IR camera image of IR transparent 20 μm thick ultra-thin 3YSZ(top) and IR opaque Corning Willow glass (bottom))

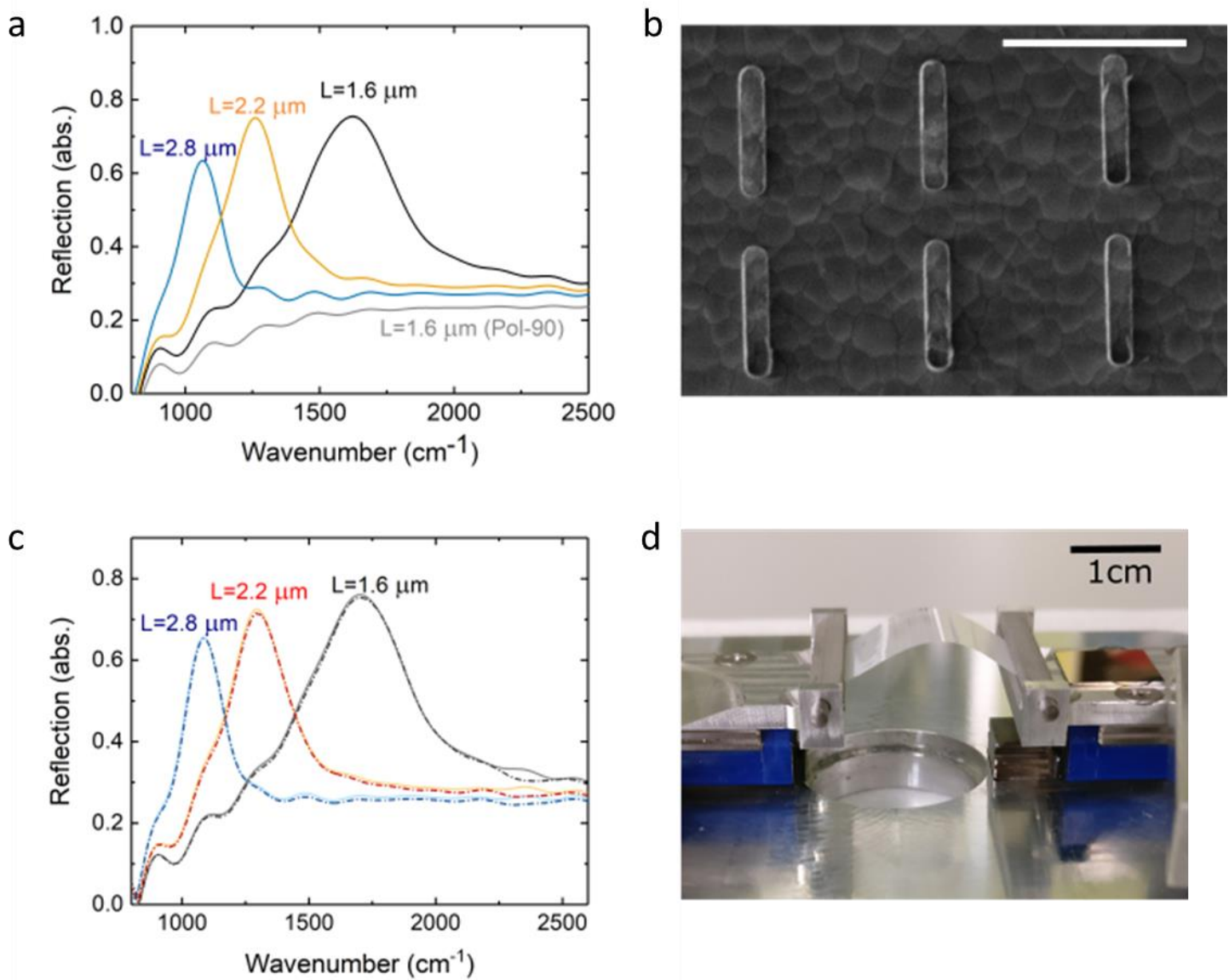

**Figure 2.** (a) Reflection spectra of gold dipole arrays of different dimensions. L is the length of the dipole and the period is 1.4 L. (b) SEM of one such gold dipole array. The scale bar is 2 µm. (c) Reflection spectra of Au dipole arrays of three different dimensions before and after (dashed line) more than 100 bending cycles. (d) Set up used to bend the 3YSZ substrate with a bending radius of 2.2 cm.

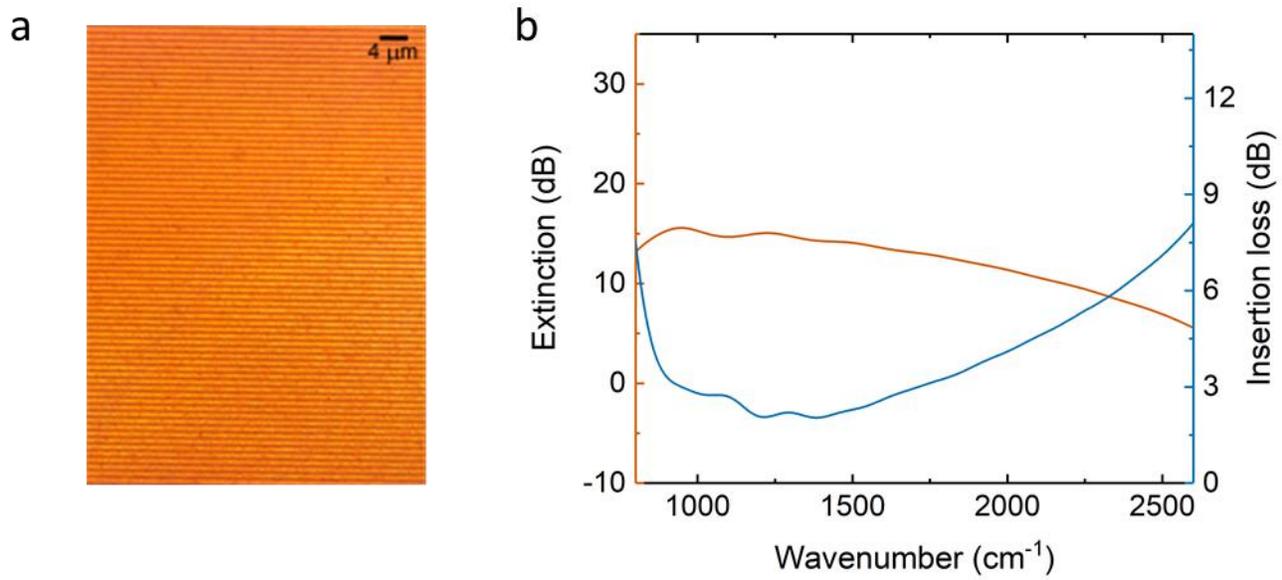

**Figure 3.** (a) Optical microscope image of wire grid polarizer. (c)Insertion loss (maximum transmission) and Extinction of a wire grid polarizer on 3YSZ.

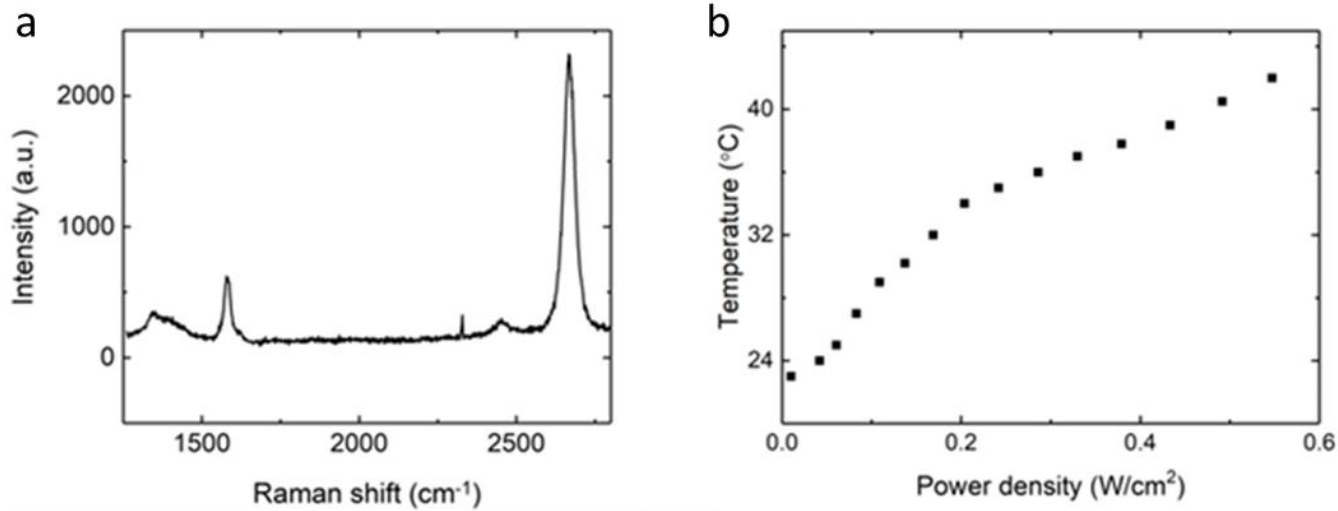

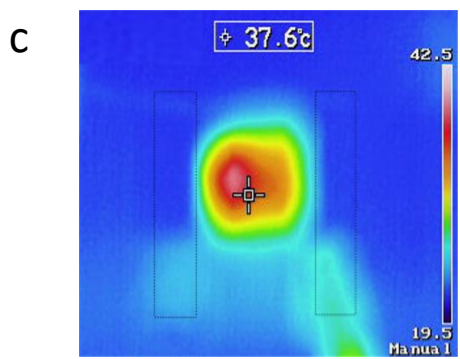

**Figure 4.** (a)Raman spectrum of graphene on 3YSZ (b) Temperature as a function of current in graphene (c) IR camera image of graphene joule heater on 3YSZ.